\documentclass[12pt,preprint]{aastex}
\usepackage{graphicx}
\usepackage{amssymb}
\usepackage{natbib}
\usepackage{amsmath}

\newcommand{\degree}{^{\circ}}

\begin{document}

\title{Influence of the ambient solar wind flow on the propagation behavior of interplanetary CMEs}

\author{Manuela Temmer$^{1}$, Tanja Rollett$^{1,2}$, Christian M\"ostl$^{1,2}$, Astrid~M. Veronig$^{1}$, Bojan Vr\v{s}nak$^{3}$, Dusan Odstr\v{c}il$^{4}$}
\affil{Kanzelh\"ohe Observatory-IGAM, Institute of Physics, University of Graz,
Universit\"atsplatz 5, A-8010 Graz, Austria}

\affil{Space Research Institute, Austrian Academy of Sciences,
Schmiedlstrasse 6, A-8010 Graz, Austria}

\affil{Hvar Observatory, Faculty of Geodesy, University of Zagreb,
Ka\v{c}i\'{c}eva 26, HR-10000 Zagreb, Croatia}

\affil{Cooperative Institute for Research in Environmental Sciences,
University of Colorado at Boulder, Boulder, Colorado, USA}

\begin{abstract}
 We study three CME/ICME events (2008 June 1--6, 2009 February 13--18, 2010 April 3--5) tracked from Sun to 1~AU in remote-sensing observations of STEREO Heliospheric Imagers and in situ plasma and magnetic field measurements. We focus on the ICME propagation in IP space that is governed by two forces, the propelling Lorentz force and the drag force. We address the question at which heliospheric distance range the drag becomes dominant and the CME gets adjusted to the solar wind flow. To this aim we analyze speed differences between ICMEs and the ambient solar wind flow as function of distance. The evolution of the ambient solar wind flow is derived from ENLIL 3D MHD model runs using different solar wind models, namely Wang-Sheeley-Arge (WSA) and MHD-Around-A-Sphere (MAS). Comparing the measured CME kinematics with the solar wind models we find that the CME speed gets adjusted to the solar wind speed at very different heliospheric distances in the three events under study: from below 30~R$_\odot$, to beyond 1 AU, depending on the CME and ambient solar wind characteristics. ENLIL can be used to derive important information about the overall structure of the background solar wind, providing more reliable results during times of low solar activity than during times of high solar activity. The results from this study enable us to get a better insight into the forces acting on CMEs over the IP space distance range, which is an important prerequisite in order to predict their 1~AU transit times.
\end{abstract}

\section{Introduction}

The evolution of coronal mass ejections (CMEs) is mainly governed by the Lorentz and the aerodynamic drag force. Initially, the CME is launched and driven by the Lorentz force, whereas the drag force becomes dominant in the later phase of the evolution as the CME propagates into IP space \citep{chen89,cargill96,tappin06,howard07}. In the first approximation the unit-length Lorentz force can be written as $F_{L}$=$I$$\times$$B$ where $I$ is the electric current within the erupting loop and $B$ is the magnetic field. The electric current and size of the current-carrying structure are related to the erupting magnetic flux. Assuming that the magnetic flux is preserved during the eruption due to the frozen-in condition (ideal-MHD), the electric current decreases when the structure enlarges, i.e.\ moves away from the Sun, which in turn decreases $F_{L}$ as well as the free magnetic energy contained in the system \citep[e.g.][]{jackson98,chen96,kliem06,subramanian07}. A prolonged magnetic reconnection below the eruption adds poloidal flux to the erupting structure sustaining the outward directed Lorentz force \citep{chen96} which powers and prolongs the CME acceleration \citep[][]{lin00,vrsnak08}. As soon as the drag force becomes dominant the CME speed will decrease until it becomes adjusted to the ambient solar wind speed \citep[][and references therein]{chen96,gopal00,vrsnak04CME,cargill04}. In its simplest form the drag acceleration can be expressed as $a_{D}=\pm\gamma|v-w|^{\alpha}$ with $\alpha$=[1,2], $w$ the solar wind speed, $v$ the CME speed, and $\gamma$ the drag parameter \citep[cf.][]{vrsnak02gopal}, the acceleration being positive if $v>w$ and negative for $v<w$. From coronagraphic observations it is obtained that a significant fraction of fast CMEs starts to decelerate already in the high corona \citep{stcyr99,vrsnak04CME,davis10}. Interplanetary scintillation was one of the earliest techniques to investigate the solar wind in the inner heliosphere \citep[see, e.g.,][]{hewish64,houminer72}. First insight into the heliospheric distance range with respect to CMEs was derived from HELIOS spacecraft \citep[e.g.][]{jackson85}. Studies using radio and scintillation measurements could gain deeper knowledge on the evolution of interplanetary CMEs, so-called ICMEs \citep[e.g.][and references therein]{manoharan00,manoharan06,reiner07}. Since 2003, data from Coriolis/SMEI using interplanetary scintillation methods reveal more details on ICMEs \citep[e.g.][]{webb06}. The STEREO mission enables us since 2006 to follow CMEs using direct imaging for the entire propagation distance from Sun to Earth and to systematically study ICMEs. First studies on the solar wind drag using STEREO data were made by \cite{byrne10} and \cite{maloney10}. However, the heliospheric distance at which the drag force finally prevails over the magnetic driving force is still not known mainly due to the unknown solar wind speed distribution in interplanetary (IP) space. The determination of the drag force is crucial in order to reliably represent the evolution of CMEs in the heliosphere and to predict its transit time to 1~AU and, thus, its possible impact at Earth \citep[e.g.][]{vrsnak02gopal,owens04,vrsnak07zic,morrill09,webb09,vrsnak10}.

The main parameters determining the drag force $a_{D}$ are speed, mass, and size of ICMEs as well as speed and density of the ambient solar wind flow \citep[see e.g.][]{vrsnak04CME,vrsnak10}. Based on the simple expression for $a_{D}$ we focus in this study on the speed differences between the ICME and the ambient solar wind flow. As we would like to know at which distance range from the Sun the CME gets adjusted to the solar wind flow and how this affects the observed CME propagation, we need to derive the solar wind speed distribution as a function of distance and time. The ambient solar wind properties are usually estimated from in situ measurements at 1~AU, however, this does not reflect their spatial distribution in IP space. As empirical approximation of how the solar wind is structured, the relation between coronal hole areas/location on the Sun and solar wind speed can be used \citep{temmer07CH,vrsnak07SW}. Again, we face the problem that only the behavior of the solar wind at the boundaries, Sun and Earth, are known but not its distribution in between. Applying numerical MHD modeling we may overcome this limitation. Significant progress has been made in current tools like ENLIL \citep{odstrcil99,odstrcil03} which allows the simulation of the solar wind conditions for the entire Sun-Earth distance based on photospheric magnetogram input and potential field source surface extrapolation. Thus, ENLIL enables us to infer the distribution of solar wind parameters in IP space and will be used to study the environmental conditions in which the CME is embedded in.

\begin{figure*}
\epsscale{1}
 \plotone{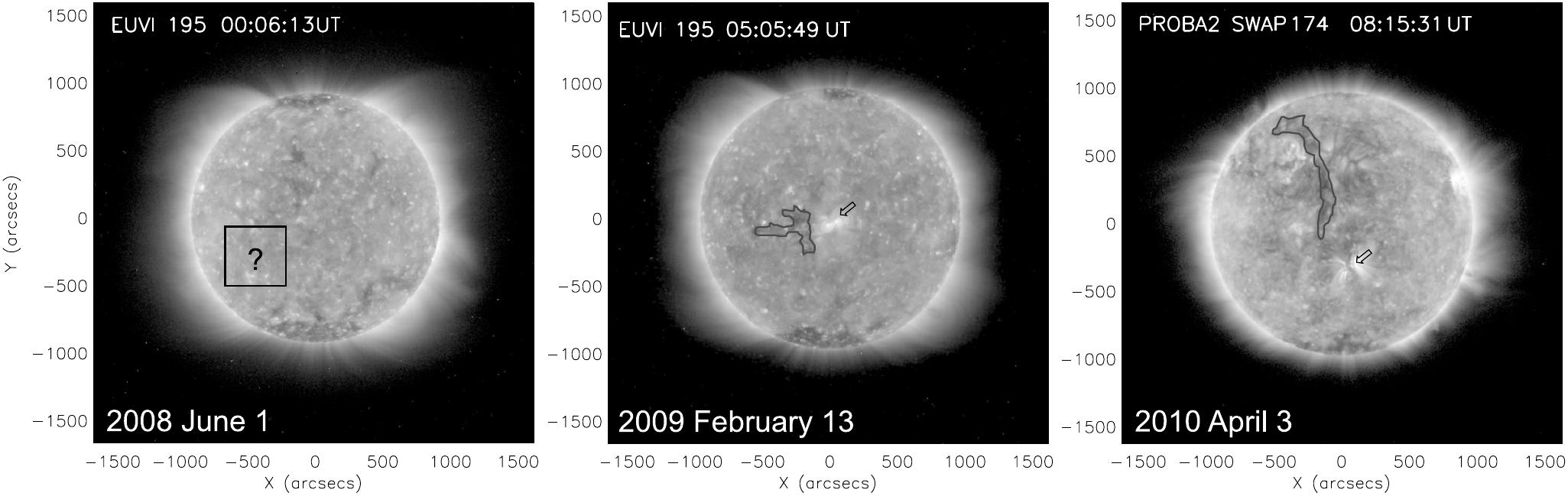}
 \caption{EUV observations from STEREO-B/EUVI in the wavelength range 195~\AA~and PROBA2/SWAP in 174~\AA~showing the solar corona at launch date for each event under study. Coronal holes in the vicinity of the source region of the CME event are outlined by solid gray lines. For Event~1 no solar surface signature is observed; the CME most likely started from the South-East quadrant as derived in the study by \cite{robbrecht09}. }
    \label{chs}
\end{figure*}

\begin{figure*}
\epsscale{1}
 \plotone{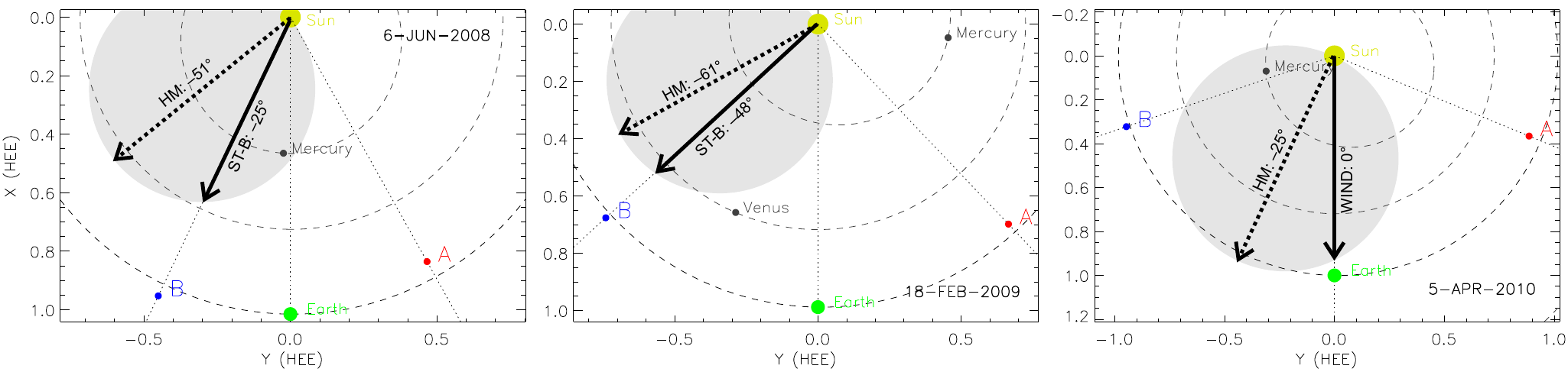}
 \caption{Location of the STEREO-A and -B spacecraft with respect to Earth for the three events under study. Assuming the CME is a circle (gray shaded) attached to the Sun, the apex of the CME (dashed arrow) is at a different distance than the flanks (solid arrow) directed towards the in situ spacecraft. The ``corrected'' ICME kinematics we discuss with respect to the background solar wind are all extracted along the Sun-spacecraft lines.}
    \label{all}
\end{figure*}

The subject of the current study is to infer the heliospheric distance at which the drag force starts to prevail over the driving force, both acting on the ICME until the speed of the ICME gets finally adjusted to the speed of the ambient solar wind flow. To this aim we study the evolution of three well observed CME/ICME events tracked all the way from Sun to 1~AU from STEREO/SECCHI remote sensing observations. In combination with in-situ measurements at 1~AU we are able to determine the direction and speed of a CME. A possible driving is derived by measuring the kinematics of the front (sheath) of the CME under the assumption that, if the body of the CME is accelerated (by one of the two main drivers) the sheath will respond and will also move faster. To what extent the Lorentz force might contribute to driving is derived from solar surface observations (growing post-flare loops due to ongoing magnetic reconnection processes adding poloidal flux to the magnetic structure of the CME body). The contribution to driving due to high speed solar wind streams is derived from comparison of the CME/ICME speed and the solar wind speed derived from ENLIL (NASA/CCMC) model runs for the ambient solar wind flow (both as function of distance along the CME propagation direction). In contrast to previous studies analyzing the effect of the ambient solar wind by simulating the propagation of CMEs/ICMEs with MHD models \citep[e.g.][]{webb09,case08}, we approach this issue by comparing the numerically calculated background solar wind speed from ENLIL model runs with observed CME kinematics in IP space. This study is aimed to be a step in gaining deeper insight into the effect of the drag force influencing the CME/ICME propagation.

\section{Data and Methods}

All three CME/ICME events are observed with the Solar Terrestrial Relations Observatory (STEREO-A and STEREO-B) SECCHI instrument suite \citep{howard-stereo08}. SECCHI consists of two coronagraphs, COR1 and COR2, covering a plane-of-sky (POS) distance range up to $\sim$15~R$_\odot$ and the heliospheric imagers (HI), HI1 and HI2, for distances $>$15~R$_\odot$. This instrument combination allows us to track CME/ICME events in the inner heliosphere from $\sim$2~R$_\odot$ to 1~AU. All events were remotely observed from STEREO-A and appeared under this vantage point quite close to the solar limb\footnote{Therefore, we can neglect projection effects in STEREO-A COR1 and COR2 measurements.}. In situ signatures of the ejecta at 1~AU from which we can deduce its arrival time and speed are derived from solar wind magnetic field and plasma data as measured with STEREO-B/IMPACT \citep{acuna08,luhmann08}, STEREO-B/PLASTIC \citep{galvin08}, and Wind/SWE/MFI \citep{ogilvie95,lepping95}. The arrival time of the ICME at 1~AU is obtained from the sharp increase in density in front of the identified flux rope or magnetic cloud signature \cite[e.g.][]{klein82,bothmer98}. For context information on low coronal signatures of CMEs as well as on the solar surface conditions and ongoing magnetic reconnection processes (growing post-flare loops) we use observations in the EUV wavelength range from STEREO/EUVI \citep{wuelser04} and Proba2/SWAP \citep{berghmans06}. All heliographic coordinates mentioned in the paper refer to Earth view.

Figure~\ref{chs} shows the condition of the solar corona for each of the three events under study. Especially, the location and area of coronal holes in the vicinity of the launch site of the CME is of interest, since coronal holes are known to be sources of high speed solar wind streams which may influence the propagation of CMEs \citep{schwenn06,gopalswamy09CME}.

In the event of 2008 June 1 (hereinafter Event~1), the CME left the Sun at $\sim$21~UT and arrived on 2008 June 6 at $\sim$22:30 at the spacecraft STEREO-B which measured clear signatures of a large scale magnetic flux rope (we refer here to a flux rope rather than a magnetic cloud since not all parameters according to the definition of a magnetic cloud by \cite{burlaga81} could be observed) with the IMPACT and PLASTIC instruments. In front of the flux rope structure a sharp increase in density is observed which can be related to the leading edge of the CME as observed in remote sensing images. The impact speed of the ICME is derived as the average speed in the ICME sheath region and is $\sim$400~km~s$^{-1}$. The event was first analyzed by \cite{robbrecht09} who classified it as ``stealth CME'' having no obvious signatures of associated low-coronal activity (filament eruption, flare, dimming, EUV wave) on the Sun. The relation between white light images from HI and in situ plasma and magnetic field measurements at 1~AU was analyzed in \cite{moestl09a}. The three-dimensional morphology and the kinematics of the CME is studied by \cite{wood10}. A comprehensive analysis of that event including the solar surface condition is presented by \cite{lynch10}.

In the event of 2009 February 13 (hereinafter Event~2), the CME left the Sun at $\sim$6~UT and could be detected by STEREO-B with IMPACT and PLASTIC which registered an ICME starting 2009 February 18 at $\sim$10~UT. The impact speed of the ICME is derived as the average speed of the density enhancement in front of the flux rope and is $\sim$360~km~s$^{-1}$. A detailed study on the ICME using the STEREO-A/HI observations along with the in situ measurements from STEREO-B and Venus Express, which measured the magnetic field of the ICME at 0.72~AU, is given in M\"ostl et al. (2011; http://arxiv.org/abs/1108.0515). The event was also associated with a global coronal wave observed in EUV \citep{cohen09,kienreich09,patsourakos09}.

In the event of 2010 April 3 (hereinafter Event~3), the CME left the Sun at $\sim$9~UT. On 2010 April 5 at $\sim$8~UT a sharp increase in density related to an IP shock was detected by in situ measurements at the Wind spacecraft, followed by signatures of a fast ICME event. From this the impact speed of the ICME is derived as the average speed in the ICME sheath region and is $\sim$720~km~s$^{-1}$. This was the first fast CME/ICME event of solar cycle 24 with an average speed over the Sun-Earth distance range of $\sim$800~km~s$^{-1}$ \citep{moestl10,liu11,wood11}.

The elongation of the leading edge of a CME is measured following the intensity enhancements from jmaps constructed from sequences of STEREO-A/HI1 and HI2 difference images \citep{davies09}. The tracking of each CME is carried out in the ecliptic plane averaging over a latitudinal range of $\pm$16 pixels (which corresponds to $\pm$0.3$\degree$ for HI1 and $\pm$1.0$\degree$ for HI2). For each event the measurements were repeated five times in order to derive the mean value and standard deviation. Applying this procedure, the elongation errors are found to be in the range $\pm$0.1--0.3$\degree$ for HI1 and $\pm$0.3--0.4$\degree$ for HI2. Since not all events can be tracked equally well in the constructed jmaps, the errors differ from event to event. Usually, the conversion from elongation into radial distance from the Sun is accomplished by applying different methods assuming the CME to be either a small-scale or a very wide ejection. Using the fixed-$\phi$ (FP) method, the CME is approximated as point-like source propagating radially along a fixed trajectory of angle $\phi$ \citep{sheeley99,sheeley08a,sheeley08b,rouillard08}. Using the harmonic mean (HM) method, the CME front is approximated as circle which is attached to the Sun (i.e.\ assuming it to be a very wide object) with its apex propagating along the angle $\phi$ \citep{lugaz09,howard09}. Hence, by varying the propagation angle $\phi$ different results for radial distance and speed of the ICME are derived.

For our study we use the propagation directions and kinematics derived by using the ``corrected'' HM conversion method which is described and applied to the same CME/ICME events under study in the paper by Rollett et al. (2011; http://arxiv.org/abs/1110.0300), and shortly summarized in the following. By measuring corresponding in situ signatures of ICMEs at 1~AU, an additional data point in the distance-time as well as in the velocity-time regime is obtained. The in situ data point presents a boundary condition that restricts the range of suitable propagation angles $\phi$ used for converting elongation into radial distance \citep[cf.][]{moestl09a,moestl10}. The propagation angle $\phi$ used as input for HM to derive radial distances and speeds of a ICME from the measured elongations, that match best the observed arrival time of the CME and the speed of the ICME measured at the location of the in situ spacecraft, gives the most probable value of the direction of motion of the CME/ICME within the geometrical assumptions we use. Applying the usual HM method \citep{lugaz09,howard09} delivers kinematics corresponding to the apex of the CME since it assumes that the apex of the CME hits the in situ spacecraft which is not necessarily correct and, thus, makes a comparison with in situ signatures geometrically inconsistent. The ``corrected'' HM method derives the kinematics for that segment of the CME along the assumed circular structure that best matches the in situ spacecraft measurements at a distance of 1~AU (comparison between timing and speed). ``Corrected'' HM is therefore a first-order approach assuming simple geometry which tackles the issue in determining which part of the CME hits the in situ spacecraft.

For the direction towards the location of the in situ spacecraft we calculate the speed of each CME by performing numerical differentiation of the radial distance-time data using three-point Lagrangian interpolation. This simply results in lower speeds as would be derived for the calculated apex direction of the CME, applying the usual HM method, but provides a more reliable comparison with the in situ measured speed of the CME. Figure~\ref{all} shows for each CME under study the derived propagation direction for the apex as well as the direction towards the in situ spacecraft applying the ``corrected'' HM method. We would like to note that ``corrected'' HM is an alternative to other existing methods and we do not claim it to be more reliable than other methods.

From the elongation-time errors as listed above, we deduce the errors for the CME speed which are of the order of $\pm$30--150~km~s$^{-1}$. Furthermore, we have to take into account an error owing to the uncertainty in the direction of motion when converting the elongation into radial distance. The best match between remote observations and in situ data has to fulfill the criteria that the arrival time and speed of the remotely observed CME at 1~AU needs to be as close as possible to the impact time and speed of the ICME as derived from in situ data. Since both criteria are not fulfilled at the same time, we derive an uncertainty in the deduced direction of motion. These errors lie in the range of $\pm$3--10$\degree$ which leads to errors for the derived CME speeds of the order of $\pm$25--100~km~s$^{-1}$ (see Rollett et al. 2011; http://arxiv.org/abs/1110.0300). The errors from the conversion method are of the same order as the measurement errors. Thus, the error bars indicated in the plots show the errors resulting from the manual tracking of the CME/ICME front. We would like to note that the uncertainties from model assumptions (geometry and linear propagation) are not included in the presented errors since they are not known. However, the reliability of the ICME kinematics as derived by using the above described method is cross-checked at 1~AU with the in situ measured impact speed of the ICME as well as with the derived CME speed from COR2 observations close to the Sun.

The distribution of the background solar wind speed for the time range during which the CME propagates from Sun to 1~AU is derived using the numerical MHD modeling code ENLIL for the inner-heliosphere \citep{odstrcil99,odstrcil03} coupled with the coronal model MAS \citep[MHD-Around-A-Sphere;][]{linker99,mikic99,riley01} and the combined empirical and physics-based model WSA \citep[Wang-Sheeley-Arge;][]{arge00}, respectively. This allows the simulation of the solar wind conditions up to 1~AU based on full-rotation (over an entire Carrington rotation) synoptic magnetograms from NSO/Kitt Peak with WSA+ENLIL starting from $\sim$20~R$_\odot$ and MAS+ENLIL from $\sim$30~R$_\odot$. For clarity we would like to stress that we did not use the ENLIL+cone model which simulates the evolution of a CME. We only use the ENLIL solar wind model for our study to simulate the 3D distribution of the background solar wind in order to infer the characteristics of the environment through which the CME propagates.

From the ENLIL numerical modeling we extract the background solar wind speed along the obtained trajectories of the CMEs (assuming constant direction). The CME trajectory derived from remote-sensing measurements in the ecliptic plane displays only a small part of the actually extended CME/ICME structure. We make no assumption about the actual size of the CME since we are interested in the local variations of the solar wind along the tracked segment of the extended CME structure. In order to take into account local variations in the ambient solar wind flow that may be just as likely to affect the CME
evolution as those along the obtained trajectory, we bin the extracted solar wind speed over a sector of $\pm$10$\degree$ in longitude and $\pm$5$\degree$ in latitude along the obtained trajectory. Applying this size of binning, we believe to cover the ambient solar wind flow that is most strongly affecting that part of the CME on which we derive the CME kinematics. Binning over larger areas showed that spatial variations in the solar wind are smoothed out.

The model runs are performed at the NASA/CCMC and are available on request under \url{http://ccmc.gsfc.nasa.gov/}. The following CCMC model runs are used in the study: Manuela\_Temmer\_032910\_SH\_1 (CR~2070; MAS), Manuela\_Temmer\_050211\_SH\_2 (CR~2070; WSA), Manuela\_Temmer\_012710\_SH\_1 (CR~2080; MAS), Manuela\_Temmer\_050211\_SH\_1 (CR~2080; WSA), Manuela\_Temmer\_121510\_SH\_2 (CR~2095; MAS), and Manuela\_Temmer\_121510\_SH\_1 (CR~2095; WSA).

\section{Results}

\subsection{Event 1: 2008 June 1--6}

\begin{figure*}
\epsscale{1.0}
 \plotone{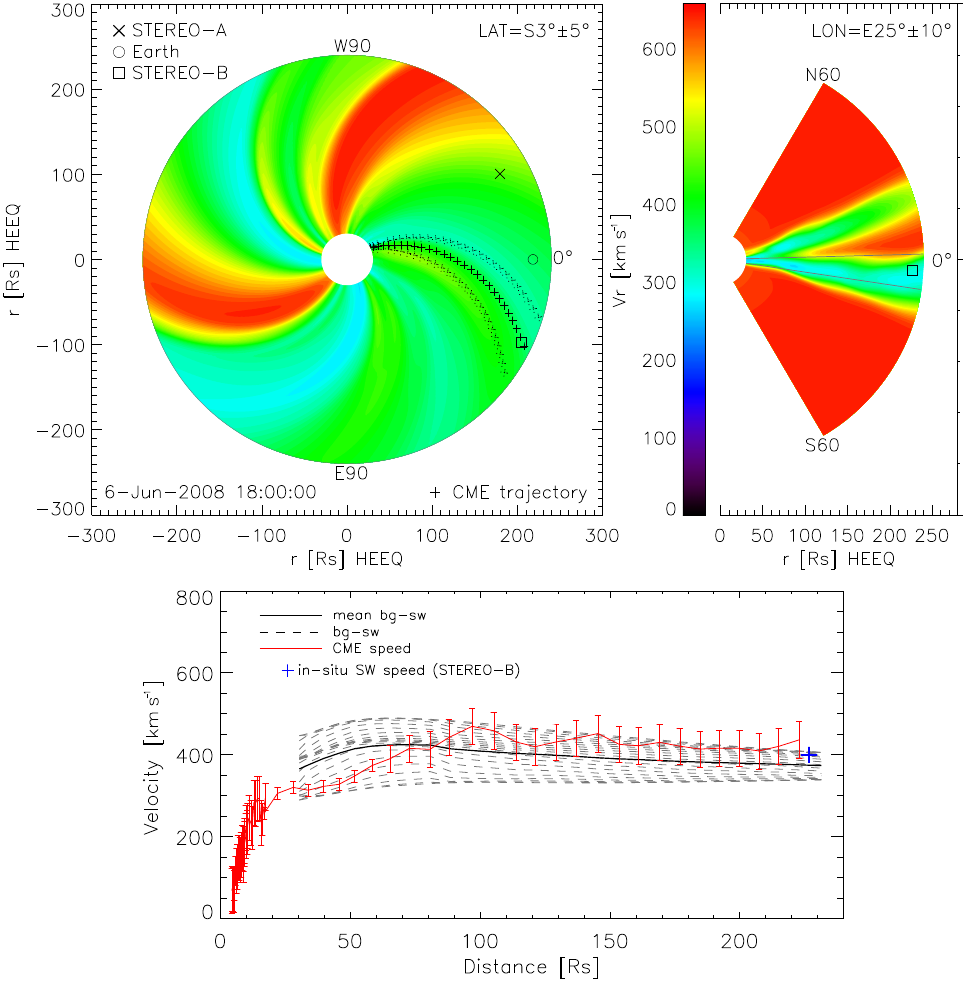}
 \caption{Event~1 - Top left: ecliptic cut (latitude S3 in HEEQ coordinates) for the background solar wind speed derived from MAS+ENLIL for CR~2070. The measured trajectory of the CME along a fixed angle of propagation at E25 (directed towards STEREO-B) is marked by black plus signs with dashed plus signs indicating a longitudinal sector of $\pm$10$\degree$. Top right: meridional cut along the direction of motion of E25. The gray lines indicate a latitudinal sector of $\pm$5$\degree$. Bottom: CME speed and errors as derived from COR and HI measurements, compared to the extracted background solar wind (bg-sw) speed rom MAS+ENLIL for $\pm$10$\degree$ (averaged over the latitudinal range of $\pm$5$\degree$) along the CME trajectory and to the in situ measured impact speed of the ICME from STEREO-B (blue cross). }
    \label{jun08-mas}
\end{figure*}

\begin{figure*}
\epsscale{1.0}
 \plotone{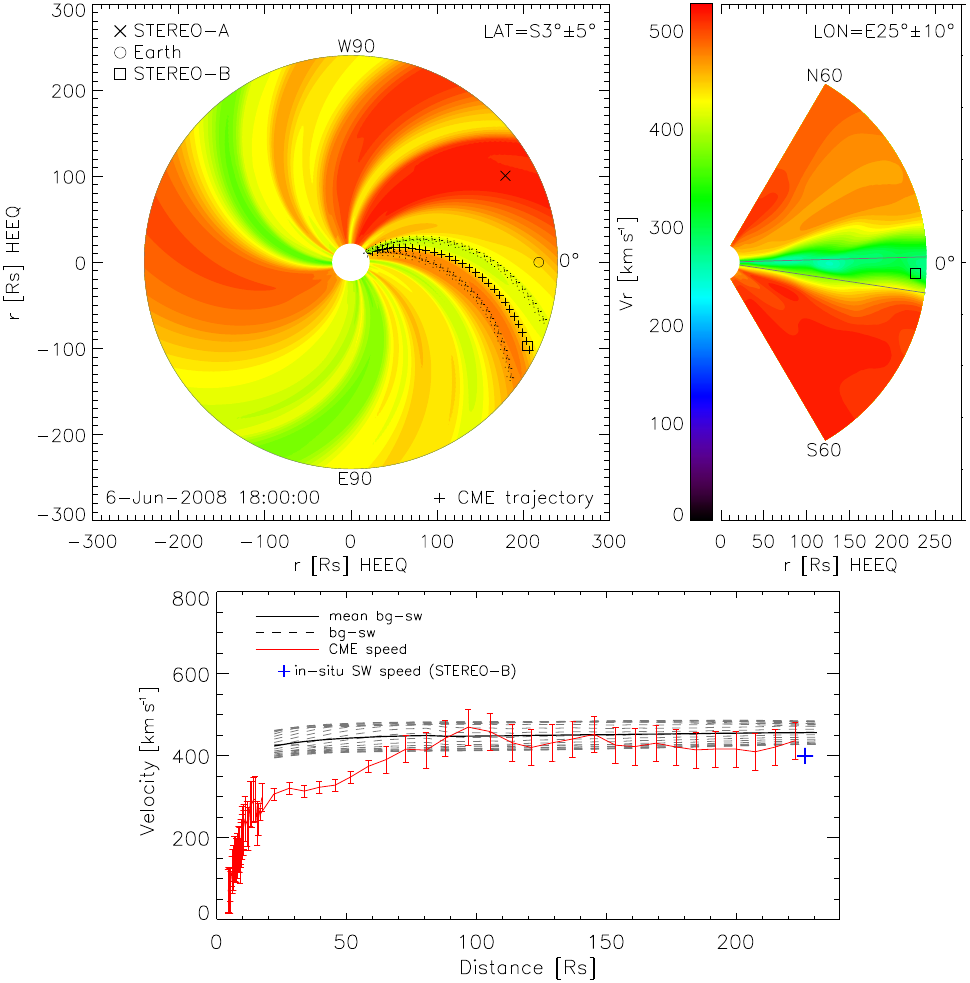}
 \caption{Same as Fig.~\ref{jun08-mas} but for WSA+ENLIL.}
    \label{jun08-wsa}
\end{figure*}

\begin{figure}
\epsscale{1.}
 \plotone{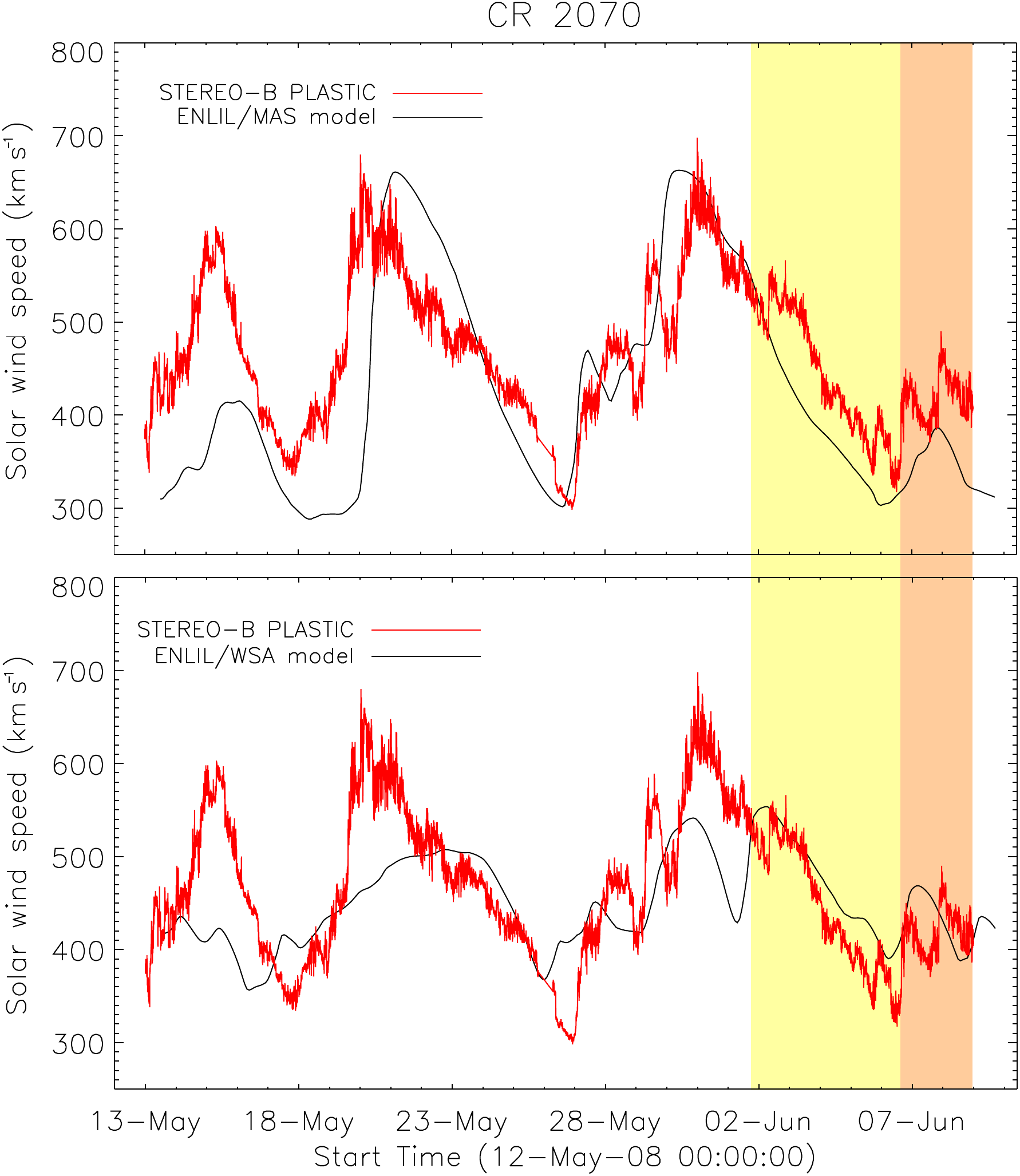}
 \caption{Comparison between in situ measured solar wind speed from STEREO-B/PLASTIC and background solar wind speed extracted from ENLIL using different coupling models for CR~2070. At the spacecraft position of STEREO-B (indicated in Fig.~\ref{jun08-mas}) the background solar wind speed is extracted from MAS+ENLIL (top panel) and WSA+ENLIL (bottom panel) over a full CR. The yellow shaded bar marks the time range of the CME event from its launch at the Sun until its in situ arrival at 1~AU. The orange shaded bar starts at the in situ arrival time of the ICME plus few days later.}
    \label{jun08-compare}
\end{figure}

Event~1 is a slow CME observed from Sun up to a distance of 1~AU with a mean speed in the COR2 FoV of $\sim$350~km~s$^{-1}$. The event is studied in detail by \cite{moestl09a} who find for the direction of motion of the CME an angle of $\sim$E35$\pm$10 (if not stated otherwise, heliographic coordinates refer to Earth view) by using the FP method as well as other reconstruction techniques which cover different distance regimes. The ``corrected'' HM method combined with in situ IMPACT and PLASTIC data from STEREO-B located at E25, gives for the apex of the CME a propagation direction of E51$\pm$6 (cf.\ left panel of Fig.~\ref{all}). Figures~\ref{jun08-mas} and \ref{jun08-wsa} (top panels)\footnote{We present the trajectory of the CME with respect to the background solar wind in such a way that the background solar wind system is kept inertial during the outward motion of the CME. Hence, we are a non-inertial observer with respect to the CME (i.e.\ positioned on the Sun) which introduces a fictitious force (Coriolis force). This causes a deviation from the expected radial motion of the CME.} show the output from MAS+ENLIL and WSA+ENLIL model runs, respectively, for Carrington Rotation (CR) 2070. A cut through the ecliptic as well as along the meridional plane of E25 gives information on how the solar wind speed is structured in IP space along the flank of the CME that hits STEREO-B. For this direction we derive the speed of the CME and extract the background solar wind speed profile from ENLIL. In the bottom panels of Figs.~\ref{jun08-mas} and \ref{jun08-wsa} we compare the observed ICME speed with the background solar wind speed from MAS+ENLIL and WSA+ENLIL, respectively, as well as with the in situ measured impact speed of the ICME at the position of the STEREO-B spacecraft. The speed of the CME derived from white light observations in the POS up to a distance of $\sim$15~R$_\odot$ can be perfectly connected to the HI speed for the distance range $>$15~R$_\odot$. Using the MAS+ENLIL model combination, the CME speed seems to be adjusted to the solar wind flow from its early evolution on whereas from WSA+ENLIL we obtain that the CME becomes adjusted to the solar wind at a distance of $\sim$70--80~R$_\odot$. As pointed out by \cite{robbrecht09} there were no signatures of magnetic reconnection in this event and concluded that the CME was not magnetically driven but rather pulled out by the solar wind. In this scenario it is not expected that the CME speed exceeds the solar wind speed in which it is embedded in. Both model results support this general picture but do not give a clear answer at which distance the CME speed gets finally adjusted to the solar wind speed.

In order to correctly interpret the results we need to evaluate the quality of the simulated background solar wind. Figure~\ref{jun08-compare} gives a comparison between the in situ measured solar wind speed from STEREO-B and the numerically calculated solar wind speed at that location. We classify the corresponding ENLIL run as reliable for our purpose if a good match is obtained between observed and numerically calculated solar wind speed during the time range covering the period between the CME launch and the in situ arrival time of the ICME plus few days later. Both models, MAS+ENLIL and WSA+ENLIL, deliver a solar wind speed which is in good agreement with the observed in situ solar wind speed (differences lie in the range of $\pm$50~km~s$^{-1}$).

\subsection{Event 2: 2009 February 13--18}

\begin{figure*}
\epsscale{1.0}
 \plotone{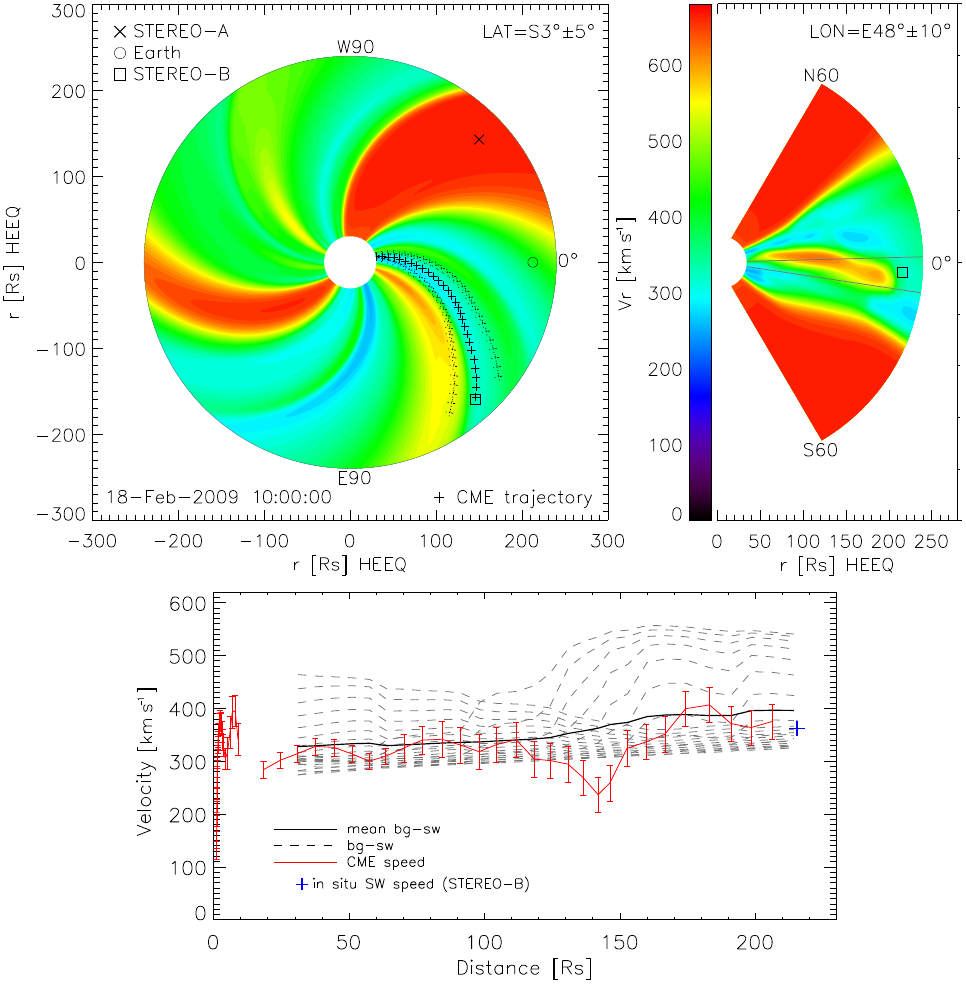}
 \caption{Event~2 - Top left: ecliptic cut (latitude S3 in HEEQ coordinates) for the background solar wind speed derived from MAS+ENLIL for CR~2080. The measured trajectory of the CME along a fixed angle of propagation at E48 (directed towards STEREO-B) is marked by black plus signs with dashed plus signs indicating a longitudinal sector of $\pm$10$\degree$. Top right: meridional cut along the direction of motion of E48. The gray lines indicate a latitudinal sector of $\pm$5$\degree$. Bottom: CME speed and errors as derived from COR and HI measurements, together with the extracted background solar wind (bg-sw) speed parameter from MAS+ENLIL (averaged over the latitudinal range of $\pm$5$\degree$) along the CME trajectory and the in situ measured impact speed of the ICME from STEREO-B.}
    \label{feb09-mas}
\end{figure*}

\begin{figure*}
\epsscale{1.0}
 \plotone{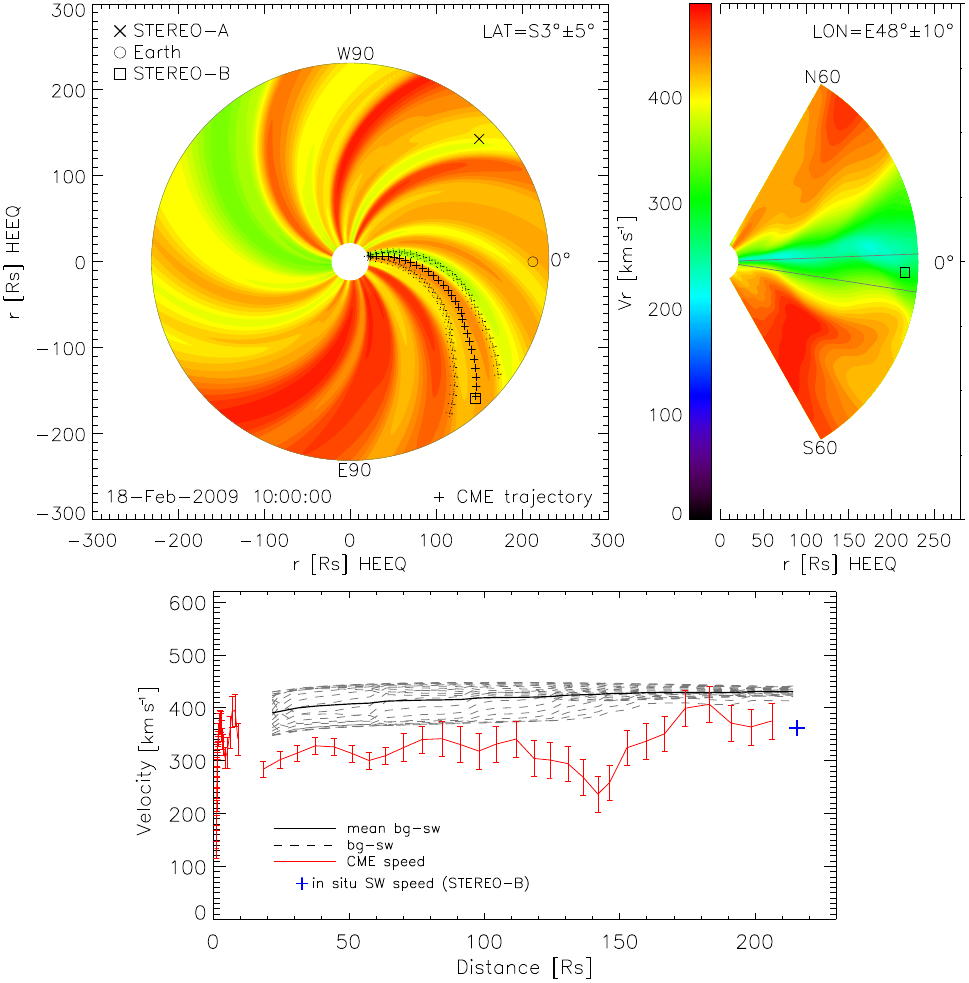}
 \caption{Same as Fig.~\ref{feb09-mas} but for WSA+ENLIL.}
    \label{feb09-wsa}
\end{figure*}

\begin{figure}
\epsscale{1.}
 \plotone{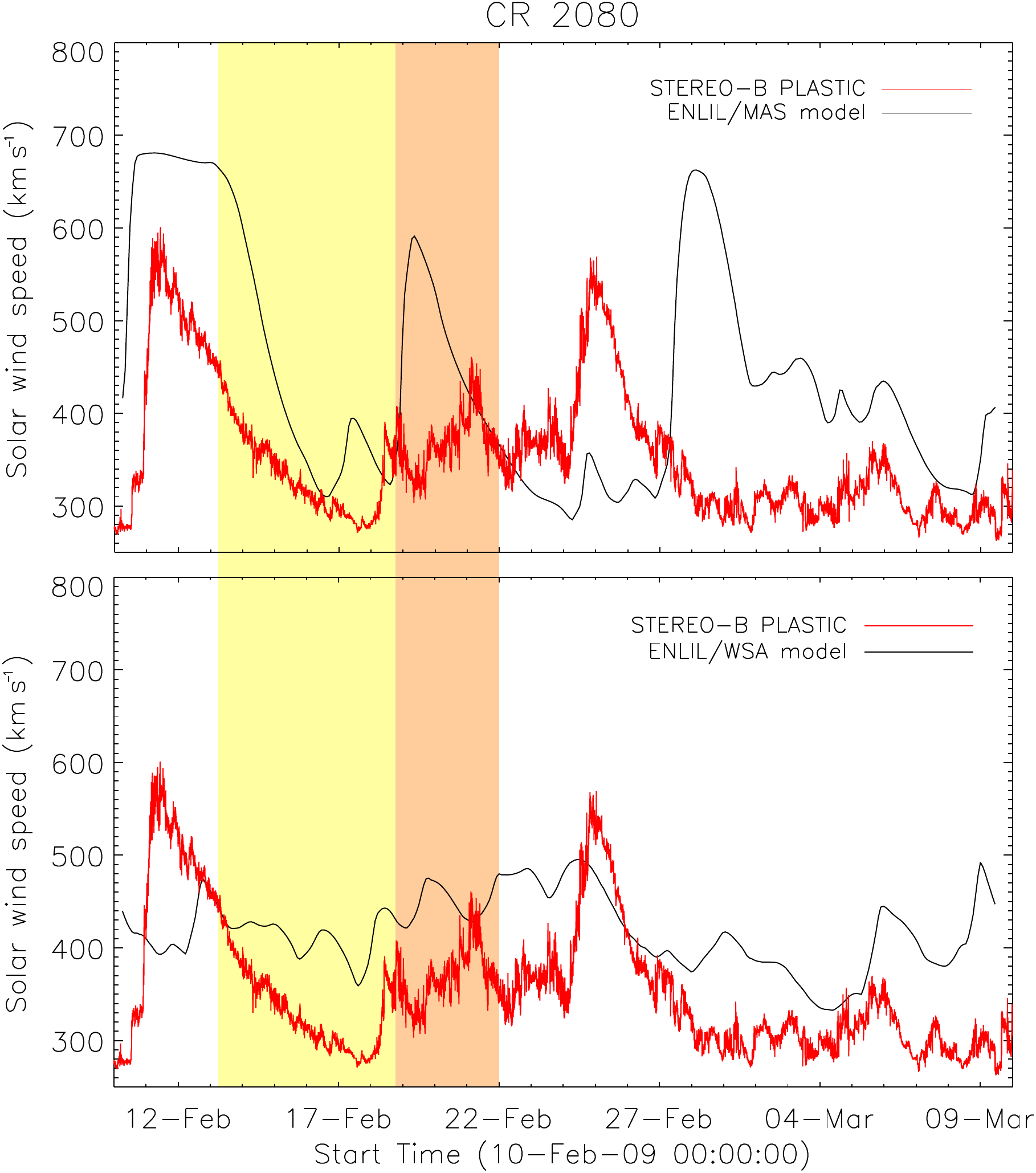}
 \caption{Same as Fig.~\ref{jun08-compare} but for Event~2.}
    \label{feb09-compare}
\end{figure}

Event~2 is a slow CME with a mean speed in the COR2 FoV of $\sim$350~km~s$^{-1}$. Applying ``corrected'' HM, the propagation direction for the apex of the CME is obtained at E61$\pm$3. This value is used to derive the speed for the direction towards STEREO-B at E48. Figures~\ref{feb09-mas} and \ref{feb09-wsa} (top panels) show the output from MAS+ENLIL and WSA+ENLIL model runs, respectively, for CR~2080 with the solar wind distribution in the ecliptic plane as well as for a meridional cut along the CME trajectory at E48. The bottom panels of Figs.~\ref{feb09-mas} and \ref{feb09-wsa} present the results for the extracted background solar wind speed along E48 compared to the CME speed derived from COR1+COR2 observations in the POS, from HI with ``corrected'' HM, and the in situ impact speed measured from STEREO-B. The CME soon becomes faint in COR2 and is tricky to follow causing a gap between COR2 and HI1 measurements leading to a difference in the derived speeds of $\sim$50~km~s$^{-1}$.

Comparing the derived CME speed to the background solar wind speed, we find from MAS+ENLIL that the CME is adjusted to the solar wind clearly below 30~R$_\odot$. From WSA+ENLIL we derive that the CME speed is smaller than the ambient medium during almost the entire propagation way from Sun to 1~AU. At a distance of $\sim$150--180~R$_\odot$ (corresponding in time to 2009 February 17--18) the observed kinematics reveals that the CME accelerates. For the time after 2009 February 17 we observe no signatures of growing post-flare loops from EUVI images, hence, no signatures of a propelling force which still accelerates the CME at this heliospheric distance. On the contrary, active region (AR) 11012 from which the CME was launched is decaying after 2009 February 14. Analyzing the distribution of the solar wind speed, both models, MAS+ENLIL and WSA+ENLIL, show an increase in the solar wind speed for that distance range. These findings suggest that the increase in the CME speed at a distance range of $\sim$150--180~R$_\odot$ is due to the increase in the ambient solar wind speed acting on the CME. Looking at the solar surface condition for that event (middle panel of Fig.~\ref{chs}), a small coronal hole is located close to the CME source region. However, the interpretation of a high speed stream from MAS+ENLIL results, extracted at the location of the STEREO-B spacecraft, is not supported by STEREO-B in situ measurements, and differences between model and observational data are of the order of $\sim$200~km~s$^{-1}$ (top panel of Fig.~\ref{feb09-compare}). Comparing model results at STEREO-B location and STEREO-B in situ measurements for the time after the arrival of the CME, WSA+ENLIL shows a better match than MAS+ENLIL (bottom panel of Fig.~\ref{feb09-compare}).

\subsection{Event 3: 2010 April 3--5}

\begin{figure*}
\epsscale{1.0}
 \plotone{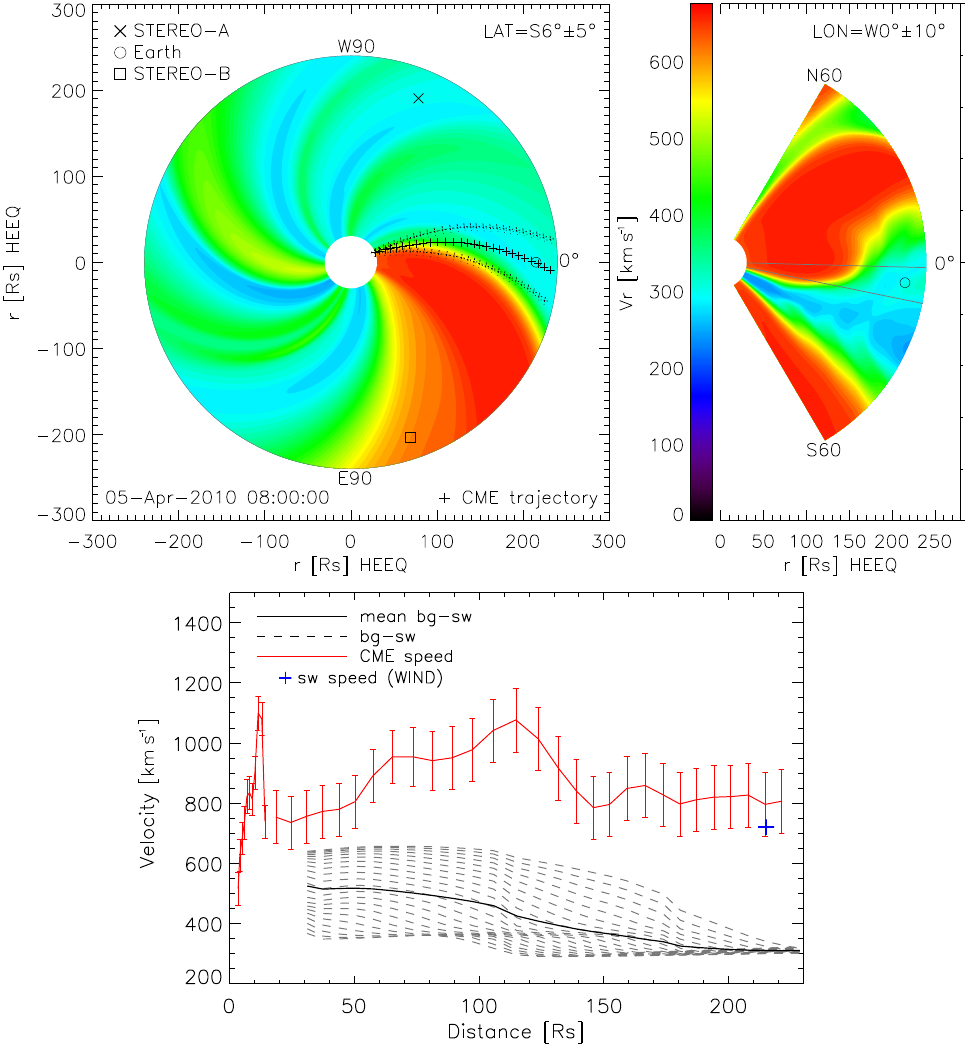}
 \caption{Event~3 - Top left: ecliptic cut (latitude S6 in HEEQ coordinates) for the background solar wind speed derived from MAS+ENLIL for CR~2095. The measured trajectory of the CME along a fixed angle of propagation at 0$\degree$ (directed towards Wind and Earth, respectively) is marked by black plus signs with dashed plus signs indicating a longitudinal sector of $\pm$10$\degree$. Top right: meridional cut along the direction of motion of 0$\degree$. The gray lines indicate a latitudinal sector of $\pm$5$\degree$. Bottom: CME speed and errors as derived from COR and HI measurements, compared to the extracted background solar wind (bg-sw) speed parameter from MAS+ENLIL (averaged over the latitudinal range of $\pm$5$\degree$) along the CME trajectory and to the in situ measured impact speed of the ICME from Wind.}
    \label{apr10-mas}
\end{figure*}

\begin{figure*}
\epsscale{1.0}
 \plotone{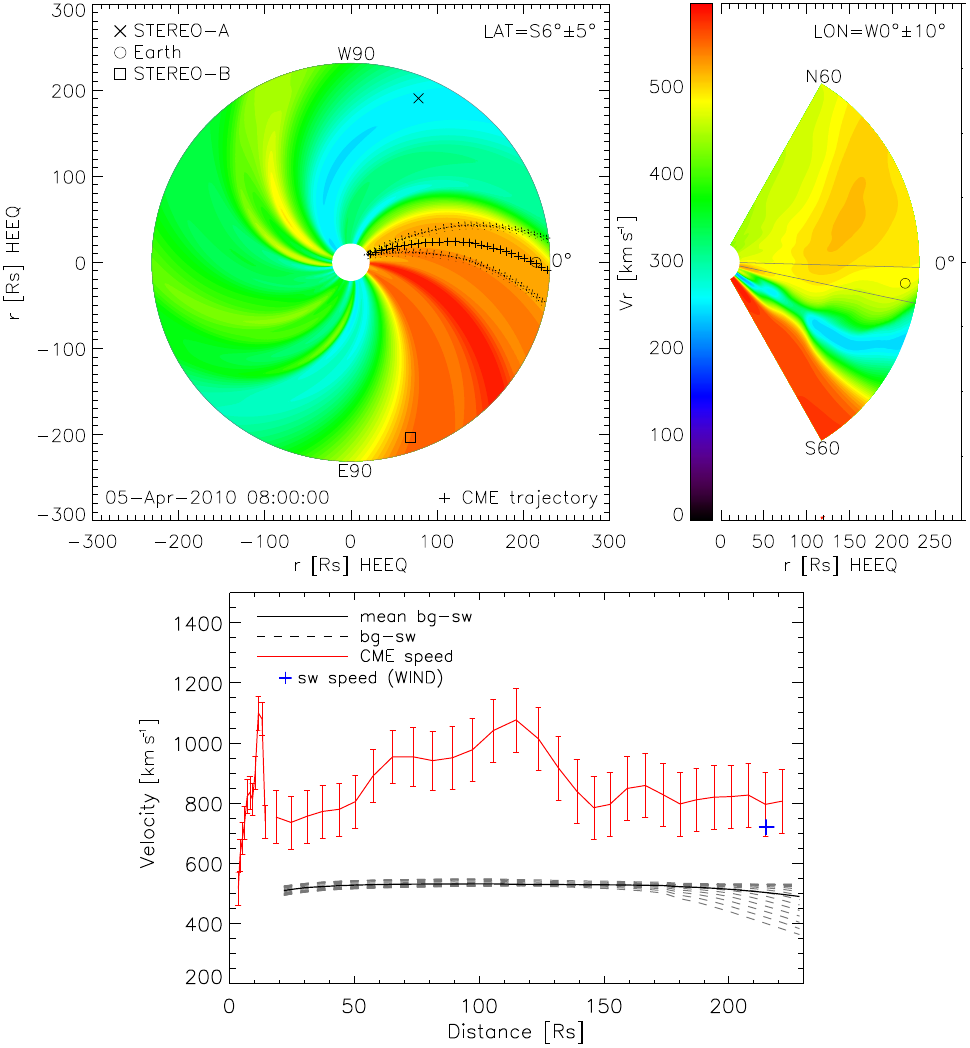}
 \caption{Same as Fig.~\ref{apr10-mas} but for WSA+ENLIL.}
    \label{apr10-wsa}
\end{figure*}

\begin{figure}
\epsscale{1.}
 \plotone{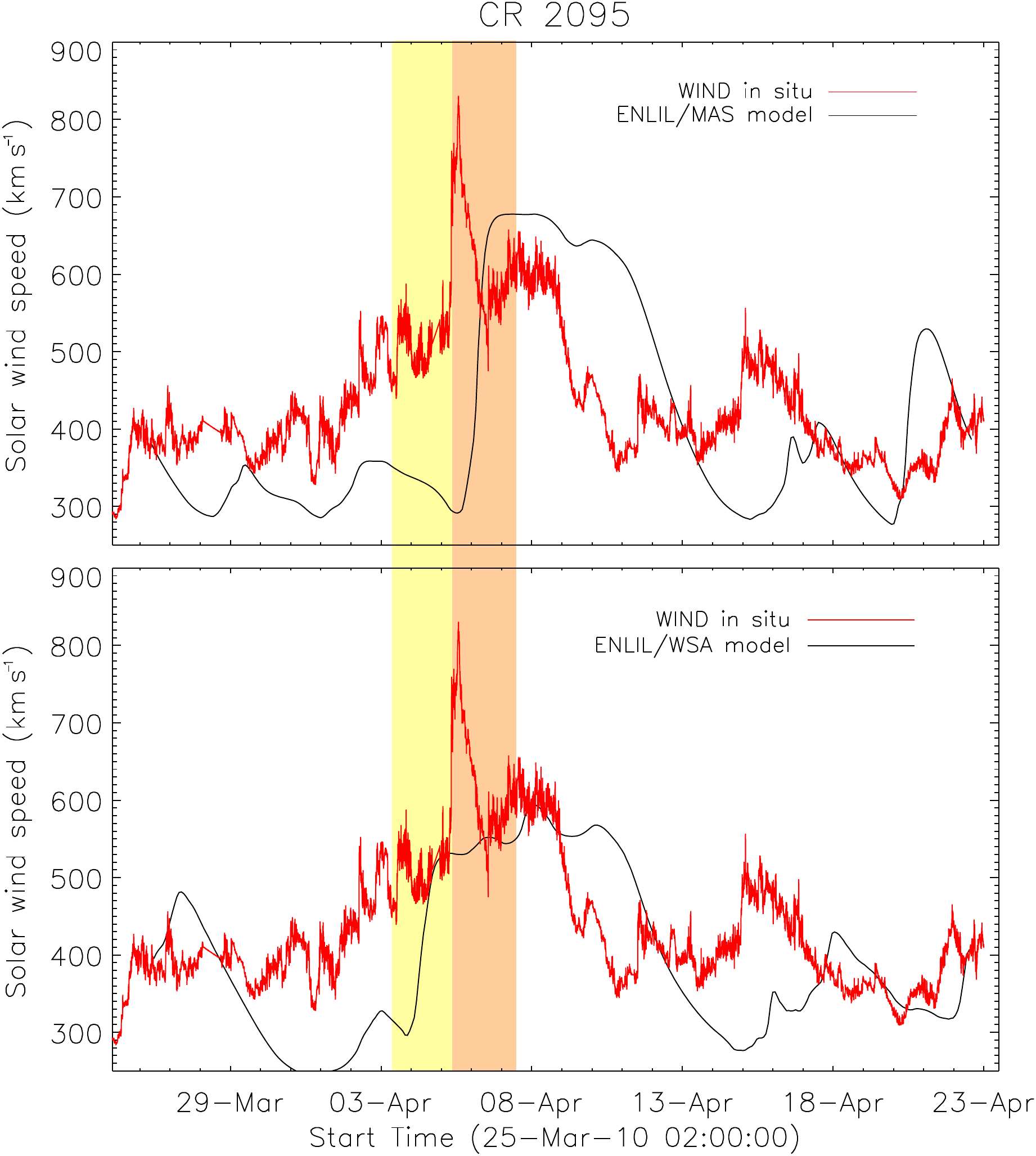}
 \caption{Same as Fig.~\ref{jun08-compare} but for Event~3 and in situ measurements from Wind. }
    \label{apr10-compare}
\end{figure}

Event~3 is a fast CME observed in situ with Wind with an average speed over the Sun-Earth distance range of $\sim$800~km~s$^{-1}$. We note that the derived trajectory of the apex of the CME of E25$\pm$10 differs by 10--30$\degree$ from the results given in \cite{moestl10} who used the usual HM method. Figures~\ref{apr10-mas} and \ref{apr10-wsa} (top panels) present the output from MAS+ENLIL and WSA+ENLIL for CR~2095, respectively, in the ecliptic and the meridional plane directed towards Earth. The bottom panels of Figs.~\ref{apr10-mas} and \ref{apr10-wsa} show the background solar wind speed extracted along the Sun-Earth line and the derived CME speed from COR1+COR2 POS observations together with the resulting HI speed of the CME using ``corrected'' HM and the in situ data point measured from Wind. The kinematics of the CME reveals a particular evolution. The CME reaches a maximum speed of $\sim$1100~km~s$^{-1}$ within the FoV of COR2 and then decelerates strongly already below 20~R$_\odot$ down to $\sim$750~km~s$^{-1}$. The final CME speed in the COR2 FoV matches the CME speed as derived from HI1 observations, from which we conclude that the strong deceleration is real. As the CME propagates within the HI2 FoV it accelerates again up to $\sim$1000~km~s$^{-1}$ and drops to a final speed of roughly 800~km~s$^{-1}$ at a distance of $\sim$150~R$_\odot$. The solar wind speed derived from both model runs, MAS+ENLIL and WSA+ENLIL, is lower than the observed ICME speed over the entire propagation path from Sun to Earth. Figure~\ref{apr10-compare} compares in situ measured and numerically calculated background solar wind speed at the location of the in situ spacecraft, revealing that none of the models reflects the solar wind speed at 1~AU for the time range of interest with differences in the range of 150--200~km~s$^{-1}$.

Several CME events with linear speeds higher than 500~km~s$^{-1}$ occurred on 2010 April 2 as listed in the CDAW catalogue \citep{yashiro04}. Coronagraphic data of the SOlar and Heliospheric Observatory (SOHO) Large Angle Spectroscopic COronagraph (LASCO; Brueckner et al., 1995) as well as imagery of the low corona in the EUV wavelength range from STEREO and SOHO, showed that all CMEs with speeds $>$500~km~s$^{-1}$, were launched from different active region(s) (northern hemisphere) than the CME under study. In addition, we checked the GOES soft X-ray flux and found no enhancement in the 1--8{\AA} channel for April 2, 2010. Therefore, we believe that the propagation of the CME under study was not significantly affected by prior events.

The derived CME kinematics vary strongly during the evolution in IP space. A coronal hole, though quite narrow, located close to the AR~11059 from which the CME is expelled, may be the source of a high speed solar wind stream (HSS). Most probably the CME crosses the HSS which influences the evolution of the CME especially of its eastern part. From both model runs a HSS is revealed but due to numerical reasons with maximum speeds of $\sim$650~km~s$^{-1}$ \citep[for details see][]{lee09}. We further note that EUVI observations from STEREO-A/B reveal growing post-flare loops of AR~1059 until 2010 April 4. Thus, the CME might be still driven up to a distance of more than $\sim$100~R$_\odot$ by the induced Lorentz force due to ongoing magnetic reconnection.

\section{Discussion and Conclusion}

Observations of three CME/ICME events tracked during their propagation from Sun to 1~AU are studied with respect to their kinematical evolution in IP space and effects resulting from the ambient solar wind. For future studies on the aerodynamic drag force owing to the solar wind, it is of special interest to know the distance at which the CME speed becomes adjusted to the ambient solar wind flow. To this aim, we applied the 3D MHD model ENLIL to simulate the steady background solar wind outflow for the inner-heliosphere and compare it with the CME speed evolution in IP space derived from STEREO-A/COR and HI observations.

In general, for each of the events under study the outcome from MAS+ENLIL and WSA+ENLIL models reveals differences in the resulting distribution of the solar wind speed over the Sun-Earth distance range, though based on the same input magnetograms (NSO/Kitt Peak). By comparing in situ measurements of the solar wind speed with the simulated solar wind speed at that location, we find that the model runs can be used to obtain a general view of the situation in IP space. For times of high solar activity both model runs give less reliable results since the occurrence of fast ejecta affecting the solar wind flow is not taken into account in the ENLIL background solar wind modeling. Most likely we find such a situation during CR~2095 covering Event~3. In a systematic comparison between model results (MAS+ENLIL and WSA+ENLIL) and in situ measurements of the solar wind parameters at 1~AU over a time range of several months, \cite{lee09} found that the overall shape and trends of the low- and high-density structures, the low- and high-speed wind streams, as well as the magnetic sector structures are replicated well within a few days. For our purpose, differences of a few days are too large since the travel time of CMEs to 1~AU is of the same order ($\sim$2--5 days).

We derive the direction of motion and speed for each CME under study by exploiting the power of combining remote sensing and in situ observations, since for all events their arrival time and plasma characteristics at 1~AU could be measured \citep[see][]{moestl09a}. By using a ``corrected'' HM method we infer the speed-distance information for that part of the CME that actually hit the spacecraft (cf.\ Rollett et al. 2011; http://arxiv.org/abs/1110.0300). For each CME we extract the numerically calculated background solar wind speed along its trajectory and compare it to the derived CME evolution. Two out of three events (Events~1 and~2) are slow CMEs with speeds of the order of 350~km~s$^{-1}$ occurring during low solar activity. Depending on the model used, MAS+ENLIL or WSA+ENLIL, we obtain quite different distance ranges at which the CME speed gets adjusted to the ambient solar wind flow.  Applying results from MAS+ENLIL for Event~1, the CME would reach the solar wind speed below 30~R$_\odot$, whereas from WSA+ENLIL at $\sim$70~R$_\odot$. According to the study of Event~1 by \cite{robbrecht09} there are no signatures of magnetic reconnection even during the very early phase of CME evolution close to the Sun. This implies that no driving forces are acting on this particular CME and that it is pulled out by the solar wind from starting from the low corona. This interpretation is supported by observations revealing a continuous increase in CME speed within the COR1+COR2 FoV matching the speed derived in the HI1 distance range. However, the inertia of the CME may cause a delay in the final adjustment. Taking into account the uncertainties in the extracted solar wind speed from WSA+ENLIL and MAS+ENLIL, the CME reaches the same speed as the ambient solar wind flow at a distance range 20--70~R$_\odot$.

From observations within the COR1+COR2 FoV, the CME speed of Event~2 clearly decelerates below 30~R$_\odot$. This can be interpreted as evidence for a strongly acting drag force over that distance range \citep[see also][]{davis10}. Results from MAS+ENLIL support this interpretation and the CME speed is most likely adjusted to the background solar wind before entering the HI1 FoV. The increase of the CME speed at a distance of $\sim$150--180~R$_\odot$ seems to be related to an increase in the background solar wind speed beyond $\sim$100--140~R$_\odot$ revealed from both model runs (MAS+ENLIL and WSA+ENLIL) rather than due to a propelling Lorentz force. This provides further evidence that the CME is well embedded in the ambient solar wind flow during its propagation in IP space.

Event~3 is the first fast CME event of cycle 24 occurring during a period of enhanced solar activity. Therefore it is more difficult to interpret from the observational as well as from the model side. The CME speed reveals a significant deceleration from $\sim$1100~km~s$^{-1}$ down to $\sim$750~km~s$^{-1}$ within the COR2 FoV and accelerates again up to $\sim$1000~km~s$^{-1}$ at a distance of $\sim$110~R$_\odot$. To explain this behavior we propose a scenario where the CME runs into strong overlying magnetic fields acting as obstacle which drastically slows down the CME already close to the Sun. Taking into account the distribution of the ambient solar wind flow on a qualitative basis, ENLIL shows that the CME crosses a HSS. Most likely we observe a very weak drag in the low-density/high-speed flow of the HSS where the Lorentz force which is still driving the CME is more effective, leading to an increase in CME speed at large distances from the Sun. As soon as the Lorentz force weakens the drag force controls the further evolution of the CME. At 1~AU the CME has a final speed of $\sim$800~km~s$^{-1}$ which is of the order of the maximum speed reported for HSSs \citep{schwenn90}. In this particular event, the final adjustment of the CME speed to the ambient solar wind flow appears to happen beyond 1~AU.

The various existing conversion methods which are used to derive radial distances from elongations all have limitations, and we have to consider artifacts that might arise in the resulting CME kinematics. Especially for the late propagation phase, hence for large elongations, the methods may reveal an (artificial) enhancement in speed \citep[see also][]{wood09,lugaz10}. However, we stress that an acceleration of a slow CME far in IP space may as well be a physical effect, caused by an increase in the ambient solar wind flow (e.g., due to solar wind high speed streams). Hence, the general assumption that CMEs show constant speed at large distances from the Sun may not be correct and depends on the characteristics of the CME and the background solar wind speed. This finding should be taken into account when using fitting routines for the conversion from elongation into radial distance which are based on the assumption of constant speed over the entire Sun to 1~AU range.

In combination with observations, ENLIL gives valuable information about the general structure of the background solar wind which enables us to interpret the observed CME kinematics over the IP space distance range. However, the uncertainty from the model outputs limits the significance of the results which would be needed for more accurate and quantitative studies (e.g., Lorentz versus drag force studies and refined prediction of 1~AU transit times).

\acknowledgments M.T., T.R., and C.M. acknowledge the Austrian Science Fund (FWF): V195-N16 and P20145-N16. The presented work has received funding from the European Commission's Seventh Framework Programme (FP7/2007-2013) under the grant agreement n$^{\circ}$~263252 [COMESEP]. The SECCHI data are produced by an international consortium of Naval Research Laboratory, Lockheed Martin Solar and Astrophysics Lab, and NASA Goddard Space Flight Center (USA), Rutherford Appleton Laboratory, and University of Birmingham (UK), Max-Planck-Institut f{\"u}r Sonnensystemforschung (Germany), Centre Spatiale de Liege (Belgium), Institut d'Optique Theorique et Appliquee, and Institut d'Astrophysique Spatiale (France). Simulation results have been provided by the Community Coordinated Modeling Center at Goddard Space Flight Center through their public Runs on Request system (http://ccmc.gsfc.nasa.gov). The CCMC is a multi-agency partnership between NASA, AFMC, AFOSR, AFRL, AFWA, NOAA, NSF and ONR. The ENLIL Model was developed by D.~Odstr\v{c}il at the University of Colorado at Boulder. We would like to thank Anna Chuklaki at NASA/CCMC for her assistance in the model computer runs needed.

\bibliographystyle{apj}

\end{document}